\documentclass[onecolumn,njp,showpacs,groupedaddress,superscriptaddress,nofootinbib]{revtex4-1}

\usepackage[pdftex,linkcolor=blue,citecolor=blue,urlcolor=blue,colorlinks]{hyperref}
\usepackage[dvipsnames]{xcolor}
\usepackage{graphicx,epsfig}
\usepackage{braket}
\usepackage{color}
\usepackage{cancel}
\usepackage{amsmath}
\usepackage{amsfonts}
\usepackage{fancyhdr}
\usepackage{soul}
\usepackage[toc,page]{appendix}
\usepackage{color}
\usepackage[normalem]{ulem}

\begin{document}
\title{Quantum sensing using imbalanced counter-rotating Bose--Einstein condensate modes}
\author{G. Pelegr\'i}
\affiliation{Departament de F\'isica, Universitat Aut\`onoma de Barcelona, E-08193 Bellaterra, Spain.}
\author{J. Mompart}
\affiliation{Departament de F\'isica, Universitat Aut\`onoma de Barcelona, E-08193 Bellaterra, Spain.}
\author{V. Ahufinger}
\affiliation{Departament de F\'isica, Universitat Aut\`onoma de Barcelona, E-08193 Bellaterra, Spain.}

\begin{abstract}

A quantum device for measuring two-body interactions, scalar magnetic fields and rotations is proposed using a Bose--Einstein condensate (BEC) in a ring trap. We consider an imbalanced superposition of orbital angular momentum modes with opposite winding numbers for which a rotating minimal atomic density line appears. We derive an analytical model relating the angular frequency of the minimal density line rotation to the strength of the non-linear atom-atom interactions and the difference between the populations of the counter-propagating modes. Additionally, we propose a full experimental protocol based on direct fluorescence imaging of the BEC that allows to measure all the quantities involved in the analytical model and use the system for sensing purposes.

\end{abstract}
\maketitle


\section{Introduction}

Pushing the limits of sensing technologies is one of the main challenges in modern physics, opening the door to high-precision measurements of fundamental constants as well as applications in many different areas of science. Specifically, the development of highly-sensitive compact magnetic field sensors enables from detecting extremely weak biologically relevant signals to localize geological structures or archaeological sites \cite{review_magnetometers}. In this context, superconducting quantum interference devices (SQUIDs) \cite{review_SQUIDS1,review_SQUIDS2} and atomic \cite{atomic_magnetometers,Kominis2003,Sheng2013,Baumgart2016,MitchellCold,MitchellHot,Kitching,Polzik} and nitrogen-vacancy diamonds \cite{NV1, NV2} magnetometers are the three main approaches that allow achieving, in a non-invasive way, unprecedented sensitivity to extremely small magnetic fields.

In particular, the extraordinary degree of control of ultracold atomic systems \cite{Bloch2008,llibreveronica} makes them ideal platforms for precision measurements \cite{Zhang2016}. There are basically two types of ultracold atomic magnetometers depending on whether the magnetic field drives the internal or the external degrees of freedom of the atoms. The former are typically based on the detection of the Larmor spin precession of optically pumped atoms while the latter encode the magnetic field information in the spatial density profile of the matter wave. Atomic magnetometers with Bose--Einstein condensates (BECs) have been investigated, for instance, by using stimulated Raman transitions \cite{magnetometryRaman}, probing separately the different internal states of a spinor BEC after free fall \cite{magnetometryfreefall}, or measuring the Larmor precession in a spinor BEC \cite{magnetometryLarmor1,magnetometryLarmor2,magnetometryLarmor3,magnetometryLarmor4,magnetometryLarmor5}. In the latter case, sensitivity can be increased by probing spin-squeezed states \cite{spinsqueezed}. In \cite{FeshbachMiscibility}, the possibility of taking profit of Feshbach resonances to use a two-component BEC as a magnetometer was also outlined. Ultracold atomic magnetometers based on detecting density fluctuations in a BEC due to the deformation of the trapping potential have also been demonstrated \cite{magnetometrydensity1,magnetometrydensity2,magnetometrydensity3}.
 
Ring-shaped potentials for ultracold atoms are a particularly interesting trapping geometry for quantum sensing and atomtronics \cite{atomtronics1,atomtronics2}.  Ring potentials are currently implemented by means of a variety of techniques, such as optically plugged magnetic traps \cite{ring1}, static Laguerre-Gauss Beams \cite{ring2}, painting \cite{ring3,ring4} and time-averaged potentials \cite{ring5,ring6,ringTaver} or conical refraction \cite{ring7}. In fact, persistent currents have been observed in BECs confined in annular traps \cite{persistent,QuenchSupercurrent} and it has also been shown that their physical behavior is in close analogy to that of SQUIDs  \cite{squid1,squid2,squid3,squid4,squid5,squid6b,squid6,squid7,squid8,squid9}. It has also been suggested \cite{ring5,squid5} that BECs in this trapping geometry could be used as rotation sensors, which have already been realized with superfluid $^3$He \cite{rotationHe} and have been proposed for matter waves based on the Sagnac effect \cite{Sagnac1, Sagnac2, Sagnac3, Sagnac4}.

In this article, we propose to use a BEC trapped in a two-dimensional (2D) ring potential for measuring with high sensitivity non-linear interactions, scalar magnetic fields and rotations. We consider an imbalanced superposition of counter-rotating Orbital Angular Momentum (OAM) modes, whose spatial density distribution presents a minimal line. A weak two-body interaction between the atoms of the BEC leads to a rotation of the minimal atomic density line whose angular frequency is directly related to the strength of such interactions. This phenomenon is somehow reminiscent of the propagation of gray solitons, which originate in repulsively interacting BECs due to a compensation between the kinetic and mean field interaction energies. In this case, however, the minimal density line appears for attractive, repulsive or even non-interacting BECs, and is a consequence of the interference between the counter-propagating modes that takes place due to the circular geometry of the system. 

The rest of the paper is organized as follows. In section \ref{sec1} we describe the physical system and we derive an analytical expression that accounts for the rotation of the line of minimal density. In section \ref{sec2}, we take profit of this expression to propose a full experimental protocol to measure the interaction strength, which is proportional to the $s$-wave scattering length. Far from the resonant field or with a dilute enough BEC, the relation between the scattering length and the applied magnetic field  given by Feshbach resonances could be exploited to use the system as a novel type of scalar magnetometer. We also outline the possibility of using the system as a rotation sensor. Finally, in section \ref{conclusions} we summarize the main conclusions. In appendix \ref{CoupledEquations} we derive the general equations that govern the dynamics of a BEC carrying OAM in a ring potential, and in appendix \ref{imaging} we give further details about the experimental implementation of the measurement protocol.

\section{Quantum sensing device}
\label{sec1}

\begin{figure}[h!]
\centering
\includegraphics[width=0.7\linewidth]{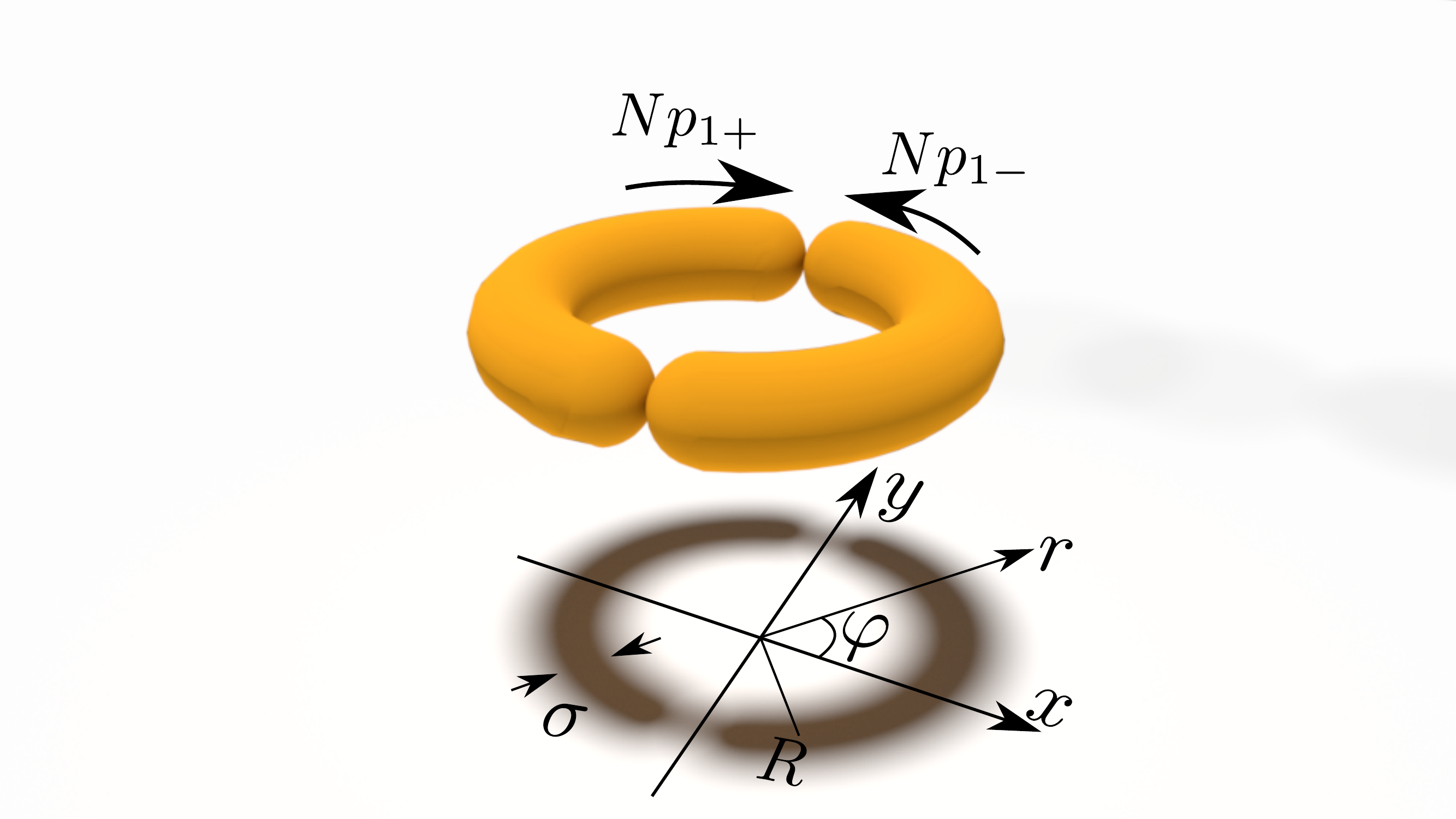}
\caption{Sketch of the physical system under consideration. A BEC formed by $N$ atoms is loaded in an annular trap, with a $p_{1+}$ population of the state $\ket{1,+}$ and $p_{1-}$ population of $\ket{1,-}$. The interference between these two counter-rotating modes yields a minimum line in the probability density. $R$ is the radius of the annulus and $\sigma$ is the width of the radial harmonic potential.}
\label{physicalsystem}
\end{figure}

\subsection{Physical system}

We consider a BEC formed by $N$ atoms of mass $m$ confined in the $z$ direction by a harmonic potential of frequency $\omega_z$ and in the perpendicular plane by an annular trap of radial frequency $\omega$ and radius $R$. We study the system in the limit of strong confinement along the $z$ direction; $\omega_z\gg \omega$. Under this assumption, in the limit $a_za_sn_2\ll 1$, where $a_s$ is the $s$-wave scattering length, $n_2$ the two-dimensional density of the BEC and $a_z=\sqrt{\hbar/(m\omega_z)}$ the harmonic oscillator length along the $z$ direction, the three-dimensional (3D) Gross--Pitaevskii equation (GPE) can be restricted to the $x-y$ plane by considering the profile for the BEC order parameter along the $z$ direction as a Gaussian of width $a_z$, which corresponds to its ground state along this direction \cite{BECbook}. In doing so, the 3D two-body interaction parameter $g_3=(N4\pi\hbar^2a_s)/m$ is transformed to its two-dimensional (2D) form $g_2=(N\sqrt{8\pi}\hbar^2a_s)/(ma_z)$ (note that in these expressions we have taken the BEC wave function to be normalized to 1). Thus, the 2D GPE that we will use to describe the system reads
\begin{equation}
i\hbar\frac{\partial \Psi}{\partial t}=\left[-\frac{\nabla^2}{{{2m}}}+V(r)+g_{2}|\Psi|^2\right]\Psi,
\label{GPE1}
\end{equation}
where $V(r)=\frac{1}{2}m\omega^2(r-R)^2$ is the potential created by the ring. Furthermore, by expressing the distances in units of $\sigma=\sqrt{\frac{\hbar}{m\omega}}$, the energies in units of $\hbar \omega$ and time in units of $1/\omega$, we arrive at the following dimensionless form of the 2D GPE, which is the one that we will use throughout the paper
\begin{equation}
i\frac{\partial \Psi}{\partial t}=H\Psi=\left[-\frac{\nabla^2}{2}+\frac{1}{2}(r-R)^2+g_{2d}|\Psi|^2\right]\Psi,
\label{GPE}
\end{equation}
where all quantities are now expressed in terms of the above defined units and the dimensionless non-linear interaction parameter is given by
\begin{equation}
g_{2d}=Na_s\sqrt{\frac{8\pi m \omega_{z}}{\hbar}}.
\label{g2d}
\end{equation}

The system supports stationary states with a well-defined total OAM $l$ and positive or negative winding number, which we denote as $\ket{l,\pm}$. The OAM eigenstates have the wave functions
\begin{equation}
\braket{\vec{r}|l,\pm}=\phi_{l\pm}(\vec{r})=\phi_{l\pm}(r,\varphi)=f_{l}(r)e^{\pm i l\varphi},
\label{OAMstates}
\end{equation}  
where $f_l(r)$ is the corresponding radial part of the wave function. 

\subsection{Dynamics in the weakly interacting regime}
Let us consider as initial state an imbalanced superposition of the $\ket{1,+}$ and $\ket{1,-}$ states, with
$n_{1\pm} \equiv p_{1+}-p_{1-}$ being the population imbalance.
Such state could be realized for instance by preparing the BEC in the ground state of the ring, imprinting a $2\pi$ round phase and momentarily breaking the cylindrical symmetry of the potential to induce a coupling between the degenerate states of positive and negative circulation \cite{reversingcirculation,geometricallyinduced} or by directly transferring OAM with a laser beam \cite{OAMLightAtoms}.

Due to parity reasons, the non-linear term in the GPE can only couple OAM states with odd total OAM $l$, see appendix \ref{CoupledEquations} for a more detailed justification. Thus, we can write the total wave function at any time $t$ as
\begin{align}
\Psi(\vec{r},t)&=\sum_{l\,{\rm odd}}\sum_{\beta=\pm} a_{l\beta}(t)\phi_{l\beta}(\vec{r}).
\label{FSM}
\end{align}
Since we focus on the weakly interacting regime, we consider that the only higher energetic states with a relevant role in the dynamics are $\ket{3,+}$ and $\ket{3,-}$. In order to simplify the forthcoming analytical expressions, we assume that the radial part of the wave functions are the ones of the ring potential ground state, i.e. we take $f_l(r)=f_0(r)$ in Eq.~\eqref{OAMstates}. This is an excellent approximation as long as the width of the the density profile of the BEC along the radial direction is much smaller than the radius of the ring, which is always the case in the weakly interacting regime. 

\begin{figure}[h!]
\centering
\includegraphics[width=0.7\linewidth]{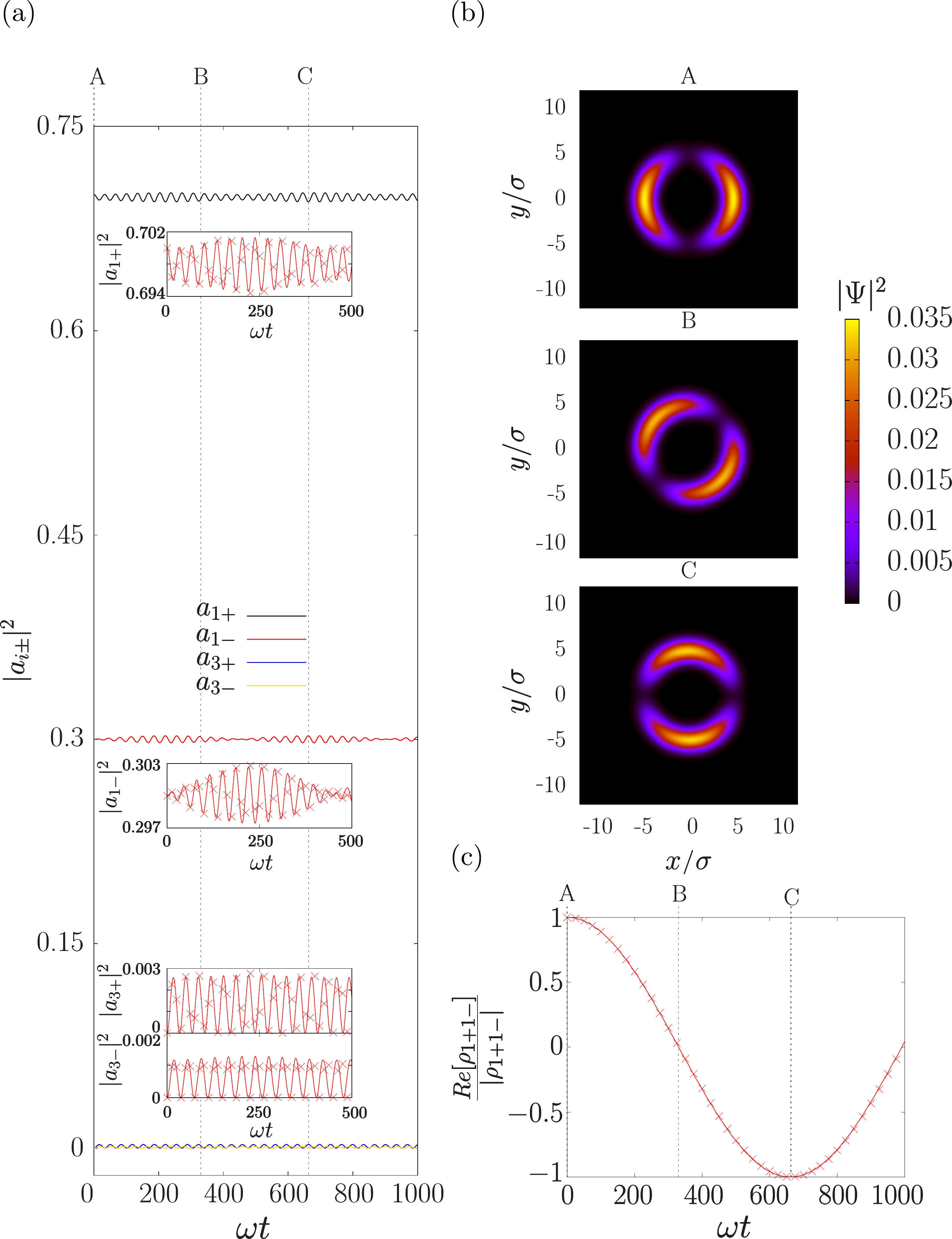}
\caption{(a) Time evolution of the population of the states involved in the dynamics. (b) Snapshots of the density profile for different instants of the dynamical evolution. (c) Time evolution of the real part of the coherence between the $\ket{1,+}$ and $\ket{1,-}$ states. The points correspond to the numerical simulation of the GPE, while the continuous lines are obtained by solving the FSM equations. The considered parameter values are $R=5$, $g_{2d}=1$, for which $U=0.0128$, ${\mu_1=0.529}$ and $\mu_3=0.699$, $a_{1+}(0)=\sqrt{p_{1+}(0)}=\sqrt{0.7}$ and $a_{1-}(0)=\sqrt{p_{1-}(0)}=\sqrt{0.3}$.}
\label{dynamics_g1_p+07_v3}
\end{figure}

The time evolution of the probability amplitudes $a_{l\pm}(t)$ ($l=1,3$) is obtained by substituting \eqref{FSM} into the GPE \eqref{GPE} (see appendix \ref{CoupledEquations} for details) 
\begin{equation}
i\frac{d}{dt}
\begin{pmatrix}
a_{1+}\\
a_{1-}\\
a_{3+}\\
a_{3-}
\end{pmatrix}
=H_{\text{FSM}}
\begin{pmatrix}
a_{1+}\\
a_{1-}\\
a_{3+}\\
a_{3-}
\end{pmatrix}.
\label{dynamicsFSM}
\end{equation} 
where the four-state model (FSM) Hamiltonian reads
\begin{widetext}
\begin{equation}
H_{\text{FSM}}
=U
\begin{pmatrix}
\mu_1/U&\rho_{1+1-}+\rho^*_{1+3+}+\rho_{1-3-}&\rho^*_{1+1-}+\rho_{1+3+}+\rho^*_{1-3-}
&\rho_{1+3-}+\rho^*_{1-3+}\\
\rho^*_{1+1-}+\rho_{1+3+}+\rho^*_{1-3-}&\mu_1/U&\rho^*_{1+3-}+\rho_{1-3+}
&\rho_{1+1-}+\rho^*_{1+3+}+\rho_{1-3-}\\
\rho_{1+1-}+\rho^*_{1+3+}+\rho_{1-3-}&\rho_{1+3-}+\rho^*_{1-3+}&\mu_3/U&\rho_{3+3-}\\
\rho^*_{1+3-}+\rho_{1-3+}&\rho^*_{1+1-}+\rho_{1+3+}+\rho^*_{1-3-}&\rho^*_{3+3-}&\mu_3/U\\
\end{pmatrix},
\label{hamFSM}
\end{equation}
\end{widetext}
where $\rho_{i\pm j\pm}\equiv a_{i\pm}a^*_{j\pm}$ with $i,j=1,3$ are the density matrix elements, $\mu_{l}$ the chemical potential of the ${l=1,3}$ OAM states, $H\phi_{l\pm}=\mu_{l}\phi_{l\pm}$, and ${U=g_{2d}\int d^2r |f_0(r)|^4}$. From these parameter definitions, the validity condition of the weakly interacting regime reads ${(\mu_3-\mu_1)\equiv \Delta \gg U}$. 
Within this regime, Fig.~\ref{dynamics_g1_p+07_v3}(a) shows a typical temporal evolution of the populations of all the OAM states involved in the dynamics considering as initial state an imbalanced superposition of the $\ket{1,+}$ and $\ket{1,-}$ states. The continuous lines have been obtained by solving with a high order Runge-Kutta method the FSM, Eq.~\eqref{dynamicsFSM}, and
the insets show the comparison with the results obtained by a full numerical integration of the 2D GPE (points). We have performed this integration using a standard Crank-Nicolson algorithm in a space-splitting scheme \cite{SpaceSplitting}, i.e., we have introduced the Trotter decomposition $e^{iH(x,y)\Delta t} \approx e^{iH(x)\Delta t} e^{iH(x)\Delta t}$, where $\Delta t$ is the discrete time step, that we have taken to be $\Delta t=10^{-3}$. The grid used for the simulations has a spatial discretization width $\Delta x=2.4\times 10^{-3}$ and a total of 1000 points in each dimension. For all the populations, we find an excellent agreement between the results obtained with the two different methods, with relative discrepancies typically on the order of $10^{-2}$. Despite the fact that the populations of the different OAM states present only very small fluctuations, the initial state is not in general a stationary state of the system because the minimum appearing in the density profile rotates at a constant speed. This fact can be appreciated in Fig.~\ref{dynamics_g1_p+07_v3}(b), where the density profile is shown for different times. At $t=0$, the density profile has a minimum density line at $x=0$, and as time marches on this line rotates in the $x-y$ plane. The fact that the minimum density line rotates means that there is a time-dependent relative phase $\alpha(t)$ between the $a_{1+}(t)$ and $a_{1-}(t)$ coefficients, so that the state of the system evolves in time as $\Psi({\vec{r},t})\approx a_{1+}(0)\phi_{1+}(\vec{r})+a_{1-}(0)e^{i\alpha (t)}\phi_{1-}(\vec{r})$. This phase difference is due to the non-linear interaction, and can be understood as a consequence of the presence of off-diagonal terms in the FSM Hamiltonian \eqref{hamFSM}. In order to determine the time dependence of $\alpha$, in Fig.~\ref{dynamics_g1_p+07_v3}(c) we plot the temporal evolution of the real part of the coherence $\rho_{1+1-}=a_{1+}(t)a^*_{1-}(t)$. We observe that it oscillates harmonically, which means that $\alpha$ evolves linearly with time. The oscillation frequency of the coherence corresponds to the rotation frequency of the minimum density line.

From the FSM, we can obtain the oscillation frequency of $\rho_{1+1-}$ by solving the {von Neumann} equation $i\dot{\rho}=[H_{\text{FSM}},\rho]$. After assuming $\rho_{1+1+}=p_{1+}$ and $\rho_{1-1-}=p_{1-}$ to be constant and neglecting all terms $\mathcal{O}(a^2_{3\pm}(t))$, we arrive at a linear system of three coupled differential equations
\begin{subequations}
\begin{align}
i\frac{d\rho_{1+1-}}{dt}&=Up_{1-}(2\rho^*_{1+3+}+\rho_{1+1-}+\rho_{1-3-})\nonumber\\
&-Up_{1+}(\rho^*_{1+3+}+\rho_{1+1-}+2\rho_{1-3-})\label{eqcoherences1}\\
i\frac{d\rho^*_{1+3+}}{dt}&=Up_{1+}(\rho^*_{1+3+}+\rho_{1+1-}+2\rho_{1-3-})+\Delta\rho^*_{1+3+}\label{eqcoherences2}\\
i\frac{d\rho_{1-3-}}{dt}&=-Up_{1-}(2\rho^*_{1+3+}+\rho_{1+1-}+\rho_{1-3-})-\Delta\rho_{1-3-}.\label{eqcoherences3}
\end{align}
\label{eqscoherences}
\end{subequations}
The characteristic frequencies $k$ of the system of equations \eqref{eqscoherences} are obtained by solving the eigenvalue equation
\begin{equation}
ik^3+ik(U\Delta+\Delta^2-p_{1+}p_{1-}U^2)+U\Delta^2(p_{1+}-p_{1-})=0.
\label{eqfreqs}
\end{equation}
Since $U \ll \Delta$ in the weakly interacting regime, the term proportional to $p_{1+}p_{1-}U^2$ can be neglected in front of the others.  The three eigenvalues that are obtained after solving Eq.~\eqref{eqfreqs} are imaginary. The eigenmode associated to the eigenvalue of lowest modulus $k_0$ has a predominant component of $\rho_{1+1-}(t)$, allowing us to write ${\rho_{1+1-}(t)\approx \rho_{1+1-}(0)e^{k_0t}}$. Thus, the rotation frequency of the nodal line is $\Omega_{\text{FSM}}=-\frac{i}{2}k_0$, where the subscript indicates that the rotation frequency has been obtained in the context of the FSM. In the limit ${\Delta \gg \Omega_{FSM}}$, the rotation frequency of the nodal line is given by
\begin{equation}
\Omega_{\text{FSM}}=\frac{U n_{1\pm}}{2(1+\frac{U}{\Delta})}.
\label{omeganode}
\end{equation}
Note that, although the $l=3$ states are nearly not populated during the dynamical evolution, the parameter $\Delta$, which contains the chemical potential $\mu_3$, plays a significant role in the expression of the rotation frequency \eqref{omeganode}. Thus, these states must be taken into account for an accurate description of the dynamics of the system.

\section{Quantum sensing protocol}
\label{sec2}
\subsection{Sensing of two-body interactions}

Recalling that the parameter $U$ of the FSM Hamiltonian \eqref{hamFSM} is given by ${U=g_{2d}\int d^2r |f_0(r)|^4\equiv g_{2d}I}$ and assuming that we are in the regime of validity of the FSM, Eq.~\eqref{omeganode} allows us to express the interaction parameter $g_{2d}$ as
\begin{equation}
g_{2d}=\frac{1}{I}\frac{2\Omega}{n_{1\pm}-2\frac{\Omega}{\Delta}},
\label{eqg2d}
\end{equation}
where $\Omega$ is the observed frequency of rotation of the nodal line. The relation \eqref{eqg2d} constitutes the basis to use the physical system under consideration as a quantum sensing device. By determining the parameters appearing on the right hand side, one can infer the value of $g_{2d}$ and thus, from Eq.~\eqref{g2d}, either the $s$-wave scattering length or the number of atoms forming the BEC. 

\begin{figure}[h!]
\centering
\includegraphics[width=0.7\linewidth]{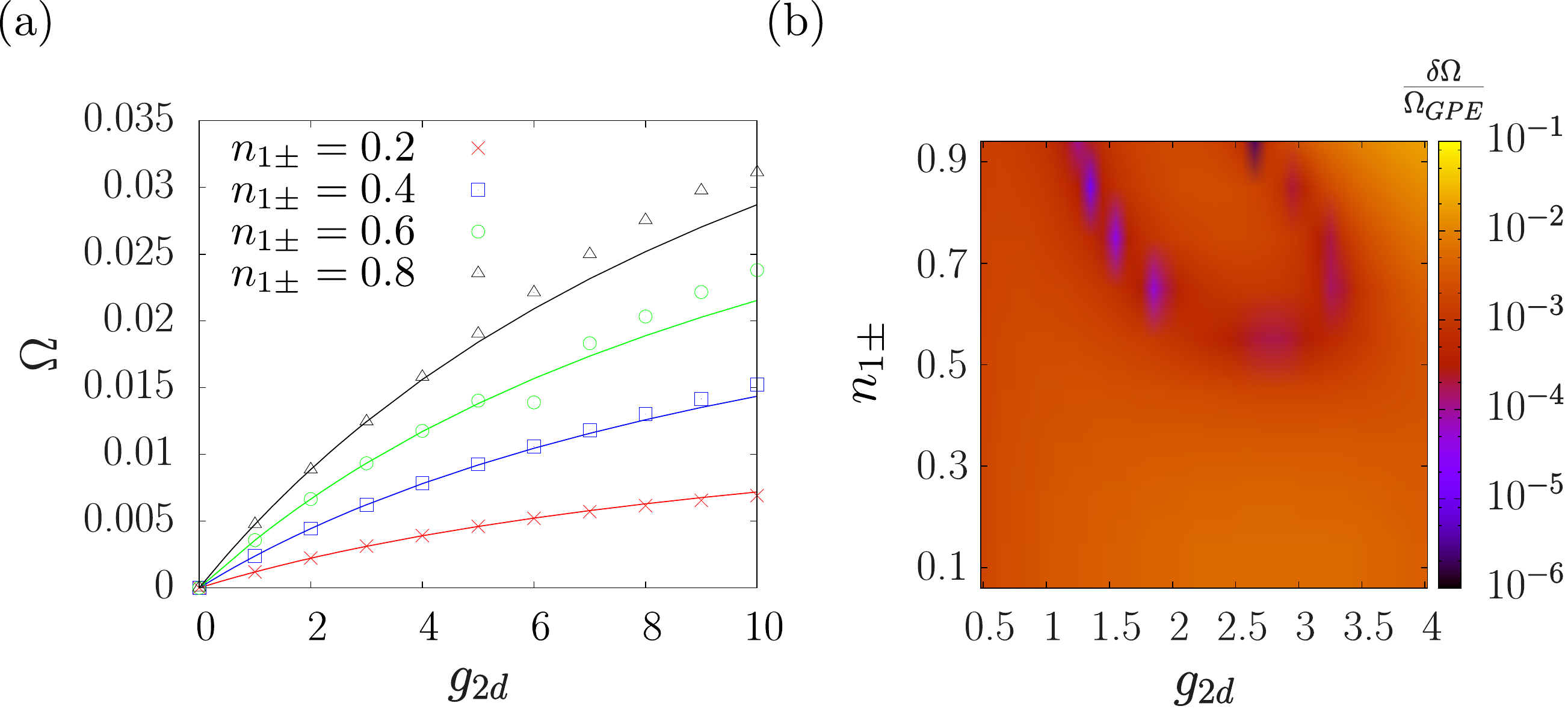}
\caption{(a) Rotation frequency of the nodal line $\Omega$ as a function of $g_{2d}$ for different values of $n_{1\pm}$ obtained with the FSM (continuous lines) and full integration of the GPE (points) (b) Relative error committed in the determination of $\Omega$ using Eq.~\eqref{omeganode} as a function of the \textit{ab initio} values of $g_{2d}$ and $n_{1\pm}$ in the simulation.}
\label{plotfreqcomplet}
\end{figure}

In Fig.~\ref{plotfreqcomplet}(a), we plot $\Omega$ as a function of $g_{2d}$ for different values of $n_{1\pm}$, computed using \eqref{omeganode} (continuous lines) and the full numerical integration of the 2D GPE (points), showing an excellent agreement between the two methods for low non-linearities and population imbalances. For $g_{2d}<4$, Fig.~\ref{plotfreqcomplet}(b) shows the relative error $\frac{\delta\Omega}{\Omega_{\text{GPE}}}$, where $\Omega_{\text{GPE}}$ is the rotation frequency of the nodal line obtained from the GPE and ${\delta\Omega=|\Omega_{\text{FSM}}-\Omega_{\text{GPE}}|}$, as a function of the \textit{ab initio} values of $n_{1\pm}$ and $g_{2d}$ in the numerical simulation, finding a maximum relative error of $10^{-2}$.
Since all the treatment developed so far is valid for low values of $g_{2d}$, this sensing device could be used for dilute BECs.

The rotation frequency of the minimum density line, $\Omega$, can be measured by direct imaging in real time of the density distribution of the BEC. If the coherence time of the BEC is $\tau$, in order for this measurement to be possible the condition $\Omega \omega \gtrsim 1/\tau$ must be fulfilled, since otherwise the rotation would be so slow that it could not be appreciated during the time that the experiment lasts. The upper limit of observable relevant values of $\Omega$ is imposed by the regime of validity of the model. If the interaction is too large, the assumptions of the FSM model are no longer valid and it is thus not possible to relate the rotation frequency of the nodal line to the non-linear interaction parameter using \eqref{eqg2d}. 
The rest of parameters appearing on the right hand side of \eqref{eqg2d} can be determined experimentally from fluorescence images of the BEC. In Appendix A we design a specific protocol to measure the population imbalance $n_{1\pm}$, the integral of the radial wave function $I$, and the chemical potential difference $\Delta$. Note that currently there are different approaches to measure the $s-$wave scattering length of ultracold atoms \cite{PethickandSmith} such as those based on photoassociation spectroscopy, ballistic expansion, and collective excitations. Our proposal constitutes an alternative to these approaches where all the unknowns can be directly inferred from fluorescence images of the BEC. However, the limit $g_{2d}<4$ obtained for the configuration discussed in Fig.~\ref{plotfreqcomplet} implies that for a BEC of, e.g.,  $10^4$ atoms of $^{23}$Na, with a trapping frequency $\omega_z$ of a few hundreds of Hz, the maximum $s-$wave scattering length that could be measured with high precision, e.g., with a relative error of $10^{-2}$, would be few times the Bohr radius.


\subsection{Sensing of magnetic fields}
Assuming that the total number of atoms of the BEC $N$ and the trapping frequency in the $z$ direction $\omega_z$ are precisely known quantities, Eqs.~\eqref{eqg2d} and~\eqref{g2d} together with the protocols to measure $n_{1\pm}$, $I$ and $\Delta$ allow to determine the scattering length $a_S$ at zero magnetic field. Alternatively, if the scattering length is a known quantity, the measurements of $\Omega$, $I$ and $\Delta$ can be used to determine $n_{1\pm}$ through the aforementioned relations.

If the scattering length depends somehow on the modulus of the external magnetic field $B$, turning on the field will be translated into a variation of $\Omega$. Thus, the system could be used as a scalar magnetometer by relating changes on the frequency of rotation of the minimal line to variations of the modulus of the magnetic field. Taking into account that $I$ and $\Delta$ are almost independent of $g_{2d}$ and thus of $B$ in the regime of interaction strengths for which the model is valid, combining Eqs. ~\eqref{g2d} and ~\eqref{omeganode} we can evaluate the sensitivity that this magnetometer would have as
\begin{equation}
\frac{d\Omega_{\text{FSM}}}{dB}=\frac{n_{1\pm} I N\sqrt{\frac{8\pi m\omega_z}{\hbar}}}{2(1+\frac{U(B)}{\Delta})^2}\frac{da_S}{dB}.
\label{dOmegadB}
\end{equation}
Since we must have $U\ll \Delta$ in order for the model to be valid, we can define a threshold limit for the sensitivity by taking $U/\Delta=1$ in \eqref{dOmegadB}. Defining the aspect ratio $\Lambda\equiv \omega_z/\omega$ and changing the differentials in \eqref{dOmegadB} by finite increments, we find the following upper threshold for the sensitivity in magnetic field variations $\Delta B_{\text{th}}$ as a function of the change in the rotation frequency of the nodal line
\begin{equation}
\Delta B_{\text{th}} =  \frac{8\sigma}{n_{1\pm} I N\sqrt{8\pi \Lambda}} \frac{1}{\frac{da_S}{dB}}\Delta \Omega.
\label{sensitivity}
\end{equation}
From Eq. ~\eqref{sensitivity}, we observe that the sensitivity is improved by having a large number of condensed particles and a strong dependence of the scattering length on the magnetic field modulus. However, since the parameter $g_{2d}\propto Na_s$ needs to be small in order for the model to be valid, it is also required that the scattering length takes small values. 

In the presence of a Feshbach resonance, the scattering length depends of the magnetic field modulus as    
\begin{equation}
a_S(B)=\tilde{a}_S\left(1-\frac{\delta}{B-B_0}\right),
\label{Feshbach}
\end{equation}
where $\tilde{a}_S$ is the background scattering length, $B_0$ is the value of $B$ at resonance and $\delta$ is the width of the resonance. Thus, by placing the magnetic field close to the resonant value $B_0$, one could in principle meet both the requirement that the scattering length is small and that it depends strongly on the magnetic field modulus. However, in most cases this procedure would have the inconvenience that close to a Feshbach resonance the three-body losses are greatly enhanced, limiting the lifetime of the BEC and hindering the measurement procedure. Nevertheless, some atomic species such as $^{85}$Rb \cite{BECRb85}, $^{133}$Cs \cite{BECCs133}, $^{39}$K \cite{BECK39} or $^7$Li \cite{BECLi7} have been reported to form BECs that are stable across Feshbach resonances, so they could be potential candidates for using the system as a magnetometer. Additionally, the BECs formed by these species have lifetimes on the order of a few seconds. Taking into account that the trapping frequency $\omega$, in units of which $\Omega$ is expressed, is typically of the order of a few hundreds of Hz for ring-shaped traps, and considering typical values of $\Omega$ shown in figure \ref{plotfreqcomplet} (a), in International System units $\Omega\sim 1$Hz. This means that in the typical time that an experiment would last, $\tau\sim 1$s, the minimum density line would perform some complete laps. Under the reasonable assumption that the fluorescence imaging system could resolve angular differences on the order of $\sim 0.1$ rad, incrementals in the rotation frequency on the order of $10^{-2}$Hz could be measured. Thus, in the dimensionless units of Eq.~\eqref{sensitivity}, sensitivites on the order of $\Delta\Omega\sim 10^{-4}$ could be achieved. These atomic species have, however, the drawback that they typically form BECs with a low number of particles, which limits the sensitivity to magnetic fields. Although it is outside of the scope of this paper to give accurate values of the sensitivities that could be achieved with this apparatus, making use of Eqs.~\eqref{sensitivity} and \eqref{Feshbach}, and considering the experimental parameters reported in \cite{squid7}, we have estimated that, in principle, this magnetometer would allow to measure changes in the magnetic field on the order of a few pT at a bandwidth of 1 Hz. 
 
As a last remark, we point out that after measuring the scattering length, far from the resonant field $B_0$, if the line of minimal density rotates at a constant speed the relation \eqref{Feshbach} can be inverted to infer the absolute value of the magnetic field. 
\subsection{Sensing of rotations}
Let us consider the case when the BEC is placed in a reference frame rotating at an angular frequency $\Omega_{\text{ext}}$, which is positive (negative) if the rotation is clockwise (counter-clockwise). Now the dynamics is governed by the modified GPE
\begin{equation}
i\frac{\partial \Psi}{\partial t}=\left[-\frac{\nabla^2}{2}+V(r)+g_{2d}|\Psi|^2-\Omega_{\text{ext}} L_z\right]\Psi,
\label{GPErotation}
\end{equation}
where $L_z=-i\frac{\partial}{\partial\varphi}$ is the $z$ component of the angular momentum operator. The ideal instance for using the system under study as a sensor of rotations is the non-interacting limit $g_{2d}=0$. In that case, it can be easily shown that the effect of the external rotation is to make the line of minimal density rotate at an angular speed $\Omega_{\text{ext}}$, which can be directly measured in experiments.

In the weakly interacting regime, the system under study can still be used as a sensor of external rotations. In that case, we find that the only difference in the dynamics with respect to the case when there is no external rotation is that the rotation frequency of the nodal line is shifted precisely by a quantity $\Omega_{\text{ext}}$. Thus, if $g_{2d}$ is known and $I$, $n_{1\pm}$ and $\Delta$ are measured using the protocol provied in the appendix A, the system under consideration can be used as a sensing device for external rotations by computing the external rotation as $\Omega_{\text{ext}}=\Omega-\Omega_{\text{FSM}}$, where $\Omega$ is the rotation frequency of the nodal line observed in the experiment and $\Omega_{\text{FSM}}$ is given by \eqref{omeganode}.

The proposed setup constitutes an alternative to the two main lines of development of rotation sensors using ultracold atoms: the atomic-gas analogues of superconducting quantum interference devices (SQUIDs) 
\cite{squid1,squid2,squid3,squid4,squid5,squid6b} and the Sagnac interferometers, for a review see \cite{Sagnac1}. Gyroscopes based on the Sagnac effect measure a rotation rate relative to an inertial reference frame, based on a rotationally induced phase shift between two paths of an interferometer and the low available atomic fluxes and low effective areas are the main limiting factors of their sensitivity.

\section{Conclusions}
\label{conclusions}
We have studied the dynamics of an imbalanced superposition of the two degenerate counter-rotating $l=1$ OAM modes of a weakly interacting BEC trapped in a 2D ring potential. We have found that the non-linear interaction induces a time-dependent phase difference between these two modes which leads to a rotation of the line of minimal atomic density of the BEC. The derived  few state model provides a simple analytical dependence between the rotation frequency and the non-linear parameter which, for low non-linearities, perfectly matches with the \textit{ab initio} numerical simulations. The measurement of the rotation frequency allows to use the system as a quantum sensor of two-body interactions, scalar magnetic fields and rotations. The theoretical treatment exposed in this work can also be extended to a regime of higher interactions, where higher OAM modes are excited and a myriad of new physical scenarios opens up.
\acknowledgements
We thank M. W. Mitchell for fruitful and stimulating discussions. We acknowledge support from the Spanish Ministry of
Economy and Competitiveness under Contract No. FIS2014-57460-P and FIS2017-86530-P, and from the Catalan Government under Contract No. SGR2014-1639 and SGR2017-1646. G.P. also acknowledges financial support from the FPI Grant No. BES-2015-073772.  

\appendix

\section{General equations of the dynamics of the OAM modes}
\label{CoupledEquations}
In this appendix, we derive the general equations that govern the dynamics of a BEC carrying OAM in the lowest vibrational state of the ring potential. These equations will allow us to justify why only states with odd values of $l$ can be excited after setting as initial state an imbalanced superposition of the $l=\pm 1$ OAM modes. We will also indicate how we have obtained the FSM Eqs. \eqref{dynamicsFSM} and \eqref{hamFSM} from the general set of equations.

We start by considering the expansion of a general state of the BEC in terms of the OAM modes
\begin{equation}
\Psi=\sum_{m}a_m(t)\phi_m(r,\varphi)=\sum_{m}a_m(t)\left[f_0(r)e^{mi\varphi}\right],
\label{ExpansionPsi}
\end{equation}
where $m\in\mathbb{Z}$ is an index that corresponds to the multiplication of the indices $l$ and $\beta$ in the expression of the OAM modes of the main text \eqref{OAMstates}. Like in the main text, we have assumed that the radial parts of the OAM modes correspond to the lowest vibrational state of the ring, i.e., $f_m(r)=f_0(r)\;\forall m$. Since the OAM wave functions are normalized to unity, $\int |\phi_m(r,\varphi)|^2rdrd\varphi=1$, the amplitudes in the expansion \eqref{ExpansionPsi} fulfill the constraint $\sum_{m}|a_m(t)|^2=1$. Substitution of the wave function \eqref{ExpansionPsi} into the 2D GPE \eqref{GPE} yields (we drop the explicit dependences on $t$ and $\vec{r}$)
\begin{align}
\sum_{l}i\frac{da_l}{dt}\phi_l&=\left[\frac{\nabla^2}{2}+\frac{1}{2}(r-R)^2+g_{2d}\sum_{m,m'}a_ma_{m'}^*\phi_m\phi_{m'}^*\right]\sum_k a_k\phi_k\nonumber\\
&=\left[\frac{\nabla^2}{2}+\frac{1}{2}(r-R)^2+g_{2d}\sum_m |a_m|^2|\phi_m|^2+g_{2d}\sum_{m\neq m'}a_ma_{m'}^*\phi_m\phi_{m'}^*\right]\sum_k a_k\phi_k\nonumber\\
&=\left[\frac{\nabla^2}{2}+\frac{1}{2}(r-R)^2+g_{2d}|f_0|^2\right]\sum_k a_k\phi_k+g_{2d}\sum_k\sum_{m\neq m'}a_ma_{m'}^*a_k\phi_m\phi_{m'}^*\phi_k
\label{EqsCoefs1}
\end{align}  
From the expression \eqref{EqsCoefs1}, an equation of motion for each of the amplitudes can be found by multiplying both sides by $\phi_l^*$ and integrating over the whole 2D space
\begin{align}
i\frac{da_l}{dt}&=\sum_k a_k\int rdrd\varphi\;\phi_l^* \left[\frac{\nabla^2}{2}+\frac{1}{2}(r-R)^2+g_{2d}|f_0|^2\right]\phi_k+g_{2d}\sum_k\sum_{m\neq m'}a_ma_{m'}^*a_k\int rdrd\varphi\;\phi_l^*\phi_m\phi_{m'}^*\phi_k\nonumber\\
&=\sum_k a_k\int rdrd\varphi\;\phi_l^* \left[\frac{\nabla^2}{2}+\frac{1}{2}(r-R)^2+g_{2d}|\phi_k|^2\right]\phi_k+g_{2d}\sum_k\sum_{m\neq m'}a_ma_{m'}^*a_k\int rdrd\varphi\;|f_0|^4e^{i\varphi (m+k-m'-l)}\nonumber\\
&=\mu_la_l+U\sum_{m\neq m'}a_ma_{m'}^*a_{(l+m'-m)},
\label{EqsCoefs2}
\end{align}
where we have defined the quantity $U\equiv g_{2d}\int rdrd\varphi\;|f_0(r)|^4$ and we have taken profit of the fact that the OAM modes $\phi_l(r,\varphi)$ are eigenstates of the time-independent 2D GPE with eigenvalue equal to their chemical potential $\mu_l$, i.e., $\left[\frac{\nabla^2}{2}+\frac{1}{2}(r-R)^2+g_{2d}|\phi_l(r,\varphi)|^2\right]\phi_l(r,\varphi)=\mu_l\phi_l(r,\varphi)$.

From Eq. \eqref{EqsCoefs2}, one can see that the term $U\sum_{m\neq m'}a_ma_{m'}^*a_{(l+m'-m)}$, which appears due to the presence of the non-linear interaction term in the GPE, introduces coupling between different OAM modes. In this paper, we have considered initial states of the form $\Psi(0)=(\sqrt{p_{1+}}\phi_1(r,\varphi)+\sqrt{p_{1-}}\phi_{-1}(r,\varphi));\; p_{1+}+p_{1-}=1$. Thus, since $a_{l}(0)=0\;\forall l\neq 1,-1$, the only higher order OAM states that will initially be directly coupled will be those in which the non-linear part of \eqref{EqsCoefs2} has a term such as $a_{1}a_{-1}^*a_{1}$ or any other combination of $l=\pm 1$ amplitudes. The only modes that have terms of this type are those of OAM $l=\pm 3$. Higher odd OAM modes, $l=\pm 5,\pm 7,...$, are subsequently populated through coupling terms that contain combinations of amplitudes of lower odd OAM modes. However, for this particular form of the initial state, the modes with an even value of the OAM, $l=0,\pm 2,...$, cannot be excited because in their dynamical equations the terms in $\sum_{m\neq m'}a_ma_{m'}^*a_{(l+m'-m)}$ always contain at least one even OAM amplitude and, since the lowest even modes $l=0,\pm 2$ are not directly coupled to the $l=\pm 1$ modes, none of the even modes will be populated during the time evolution. This justifies the expansion of the wave function in terms of only odd OAM modes of Eq. \eqref{FSM}.

The FSM Eqs. \eqref{dynamicsFSM} and \eqref{hamFSM} have been obtained directly from Eqs. \eqref{EqsCoefs2} by truncating the basis at the $l=\pm 3$ OAM modes and rewriting the resulting four non-linear coupled equations in a more compact matrix form.

\section{Fluorescence imaging protocol}
\label{imaging}

In the following lines, we describe in detail how the values of the parameters appearing in the right hand side of Eq.~\eqref{eqg2d} could be inferred experimentally by means of direct fluorescence imaging of the BEC.

\subsubsection*{1. Population imbalance}
\begin{figure}[h!]
\centering
\includegraphics[width=0.7\linewidth]{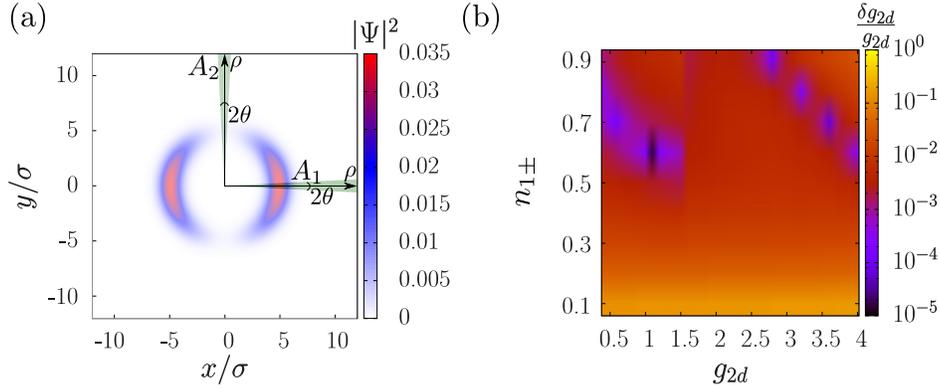}
\caption{(a) Example of $A_1$ and $A_2$ integration areas to experimentally determine the population imbalance (b) Relative error committed in the determination of $g_{2d}$ using the full experimental protocol described in the main text as a function of the \textit{ab initio} values of $g_{2d}$ and $p_{1+}$ in the simulation.}
\label{figureprotocol}
\end{figure}

The population imbalance between the two $l=1$ states can be determined from the density profile per particle at any time $t$, which can be obtained by fluorescence imaging. Since the wave function is given by $\Psi(\vec{r},t)=f_0(r)(\sqrt{p_{1+}}e^{i\varphi}+\sqrt{p_{1-}}e^{-i(\varphi+\Omega t)})$, its density profile reads
\begin{equation}
|\Psi|^2=f_0^2(r)(1+2\sqrt{p_{1+}p_{1-}}\cos 2(\varphi+\Omega t)).
\label{psisquare}
\end{equation}
Thus, the atom density has a minimum at $\varphi=\pi/2-\Omega t$ and a maximum at $\varphi=-\Omega t$. Let us now consider the two integration regions $A_1$ and $A_2$ shown in Fig.~\ref{figureprotocol}(a), which are arcs of radius $\rho$ and angle $2\theta$ centred around the maximum and minimum of intensity, respectively. The integrals of $|\Psi|^2$ over $A_1$ and $A_2$ can be performed numerically and, for sufficiently small $\theta$, they yield approximately
\begin{subequations}
\begin{align}
I_1=\int_{A_1} d^2r |\Psi|^2\approx 2\theta(1+2\sqrt{p_{1+}p_{1-}})\int_0^\rho rf_0^2(r) dr\ \label{I_1}\\
I_2=\int_{A_2} d^2r |\Psi|^2\approx 2\theta(1-2\sqrt{p_{1+}p_{1-}})\int_0^\rho rf_0^2(r) dr\ \label{I_2}
\end{align}
\end{subequations}
Thus, combining \eqref{I_1} and \eqref{I_2} one can determine the product of populations as 
\begin{equation}
p_{1+}p_{1-}=\left(\frac{I_1-I_2}{2(I_1+I_2)}\right)^2,
\end{equation}
which, together with the constraint $p_{1+}+p_{1-}=1$, allows to determine the population imbalance from a fluorescence image.
\subsubsection*{2. Integral of the radial wave function I}
From equation \eqref{psisquare}, we can write
\begin{align}
|\Psi|^4&=f_0^4(r)\left[1+4\sqrt{p_{1+}p_{1-}}\cos2(\varphi+\Omega t)\right.\nonumber\\
&\left.+4p_{1+}p_{1-}\cos^22(\varphi+\Omega t)\right].
\end{align}
From a fluorescence image, one can numerically perform the integral $\int d^2 r|\Psi|^4$ over the whole space, from which the desired quantity can be calculated as
\begin{equation}
I=\int d^2r f_0^4(r)=\frac{\int d^2 r|\Psi|^4}{1+2p_{1+}p_{1-}}.
\end{equation}

\subsubsection*{3. Chemical potential difference}
The chemical potential of the angular momentum states can be decomposed into its kinetic, potential and interaction contributions. Since one can assume that the wave functions take the form $\phi_{l\pm}(\vec{r})=f_0(r)e^{\pm il\varphi}$, the potential and interaction contributions will be the same regardless of $l$, while the kinetic contribution is given by
\begin{equation}
E^{kin}_l=\frac{1}{2}\int d^2r |\nabla \phi_{l\pm}(\vec{r})|^2=\frac{1}{2}\int d^2r \left[\left(\frac{df_0}{dr}\right)^2+l^2\left(\frac{f_0}{r}\right)^2\right].
\end{equation}
Thus, the chemical potential difference is only due to the difference in the centrifugal terms of the kinetic energy
\begin{equation}
\mu_3-\mu_1=E^{kin}_3-E^{kin}_1=4 \int d^2r \left(\frac{f_0(r)}{r}\right)^2.
\label{Delta}
\end{equation}
From Eq.~\eqref{psisquare}, one can see that the integral \eqref{Delta} can be numerically performed after determining $f_0^2(r)$ from a fluorescence image as ${f_0^2(r)=\frac{|\Psi(r,\varphi=-\Omega t,t)|^2}{1+2\sqrt{p_{1+}p_{1-}}}}$.

In order to check the accuracy of the proposed experimental protocol, we have computed $g_{2d}$ using Eq.~\eqref{eqg2d} and determining all the parameters on the right hand side  following the above described numerical procedures, and later on comparing with the \textit{ab initio} used value of $g_{2d}$ in the simulation. In Fig.~\ref{figureprotocol}(b) we plot the relative error $\frac{\delta g_{2d}}{g_{2d}}$ committed as a function of the \textit{ab initio} values of $g_{2d}$ and $p_{1+}$. In the region $g_{2d}\approx 1$ and $n_{1\pm}\approx 0.6$, the relative error is minimal and it reaches very low values, on the order of $10^{-5}$. The maximum value of the relative error is about $10\%$, and is found for low values of $n_{1\pm}$. In our simulations, we have used a grid of dimensions $24\times 24$ and 1000 points in each spatial direction. With higher grid precision, the relative error committed with the proposed protocol could prove to be even lower.

\end{document}